\documentclass[journal]{IEEEtran}
\usepackage{graphicx,subfigure}
\usepackage{amsfonts,amsmath,amssymb}
\usepackage{authblk}

\ifCLASSINFOpdf
\else
\fi
\begin{document}
\title{Diffusion Based Nanonetworking: A New Modulation Technique and Performance Analysis}
\allowdisplaybreaks

\author{Hamidreza~Arjmandi,
        Amin~Gohari,
        Masoume~Nasiri~Kenari,~\IEEEmembership{Member,~IEEE}
        and Farshid~Bateni}
\affil{Electrical Engineering Department, Sharif University of Technology}
\maketitle

\begin{abstract}
%\boldmath
In this letter, we propose a new molecular modulation scheme for nanonetworks. To evaluate the scheme we introduce a more realistic system model for  molecule dissemination and propagation processes based on the Poisson distribution. We derive the probability of error of our proposed scheme as well as the previously introduced schemes, including concentration and molecular shift keying modulations by taking into account the error propagation effect of previously decoded symbols. Since in our scheme the decoding of the current symbol does not depend on the previously transmitted and decoded symbols, we do not encounter error propagation; and so as our numerical results indicate, the proposed scheme outperforms the previously introduced schemes. We then introduce a general molecular communication system and use information theoretic tools to derive fundamental limits on its probability of error.
\end{abstract}


% \fi
\IEEEpeerreviewmaketitle

\section{Introduction}
\IEEEPARstart{N}{anonetworking} promises new solutions for many applications in the biomedical and industrial fields. This new paradigm utilizes various methods to communicate information between nano-scale machines \cite{AJ}. A promising communication method at this scale, is \emph{molecular communication}. Different molecular communication schemes have been proposed that can be categorized based on their effective range of communication, namely short, medium or large \cite{ABB}.

In this paper, we focus on a diffusion based communication system for short and medium range nanonetworks. In this system, molecule concentration represents information. The transmitter emits molecules whose type and intensity depend on the current input symbol. These molecules propagate through the environment, with a portion hitting the receiver's surface. Receptors located on receiver's surface form chemical bonds with the incoming molecules, initiating a process that eventually results in decoding of the transmitted information.

Different channel models for molecular communication have been developed and their related channel capacities have been evaluated in \cite{AA}-\cite{E}. Authors in \cite{K} have proposed two modulation methods called Concentration Shift Keying (CSK) and Molecular Shift Keying (MOSK). In these schemes,  information is encoded in molecule diffusion rate and molecule type, respectively. Both CSK and MOSK suffer from interference caused by molecules from previous transmissions, hitting the receiver after a long delay. To control this interference in CSK (where only one molecule type is used), the decoding of the current symbol is adapted to the last decoded symbol. Although this helps to manage the interference, it results in an error leakage to the current symbol if the last symbol is decoded incorrectly. Using multiple molecule types, MOSK is more resistant to interference than CSK but it requires complex molecular mechanisms at both the transmitter and the receiver for message synthesis and decoding \cite{K}.

In this paper we propose a new modulation method based on the idea of using distinct types of molecules for consecutive slots at the transmitter, thus effectively suppressing the interference. Although simple in its formulation, the method outperforms the existing ones. To evaluate the schemes, we propose a new system model for the molecule propagation process and the channel based on the Poisson distribution. This model is more realistic and amenable to analysis than the one proposed in  \cite{K} based on the binomial distribution. We evaluate and formulate the probability of error for our proposed scheme as well as the previous ones and compare the results. Specially, we take into account the dependency of decoding of the current symbol to the last transmitted and decoded symbols to compute these metrics. Numerical results indicate our scheme has a lower probability of error compared to CSK and MOSK. To understand the efficiency of our proposed scheme, we use information theoretic tools to derive a fundamental limit on the minimum probability of error of an arbitrary scheme. A comparison indicates that although our scheme outperforms the existing ones, its error exponent (in terms of transmission power) is larger than our fundamental lower bound, suggesting room for further improvement.

The rest of the paper is organized as follows. Proposed modulation scheme and system model are described in Section II. In Section III  performance of different modulation techniques are discussed. Also in this section a more general molecular communication system is introduced for which a lower bound on probability of error is computed. Lastly numerical results are presented in Section IV.

\section{Proposed Modulation Scheme and System Model}
\subsection{Previously introduced modulation schemes}

The Concentration Shift Keying (CSK) scheme is inspired by the Amplitude Shift Keying (ASK) used in the classical communication. In this scheme, symbols are encoded in the number of molecules that the transmitter diffuses; this is called \emph{diffusion rate} or \emph{transmission power}. For example, to represent $b$ bits, $2^b$ different molecule diffusion rates are utilized. At the receiver, the concentration (number of the received molecules) is compared with decision thresholds in order to decode the transmitted symbol. The molecules from the previous symbol in the last time slot may hit the receiver in the current time slot. Thus, to reduce the effect of the interference in this method, decision thresholds for the current symbol need to be adapted to the last decoded symbol.

Molecular Shift Keying (MOSK) scheme resembles the orthogonal modulation in the classical communications. To transmit $b$ bits per time slot, $2^b$ different types of molecule are utilized. The transmitter releases a specified number of molecules of the type corresponding to the current input symbol. To decode the transmitted signal at a certain time slot, the receiver looks for a unique molecule type whose concentration exceeds a certain threshold $\tau$. An error occurs if the concentration of none of the molecule types or more than one molecule type exceeds $\tau$. Although interference due to the previous transmission in MOSK is less than that in the CSK, but it still exists. Also using different molecule types complicates the hardware structure at both the transmitter and the receiver as the number of molecule types increases exponentially with $b$.
%\vspace{-0.3cm}
\subsection{Proposed modulation scheme}

The performance of both the MOSK and CSK schemes is degraded by the interference caused by molecules of previous transmissions. In addition CSK suffers from propagation of errors since the thresholds for each time slot is adapted to the previously decoded symbols. Therefore, if the previous symbol is falsely decoded, its error leaks to the current symbol.

Our scheme borrows ideas from CSK and MOSK. Having molecules of types $A_1$ and $A_2$, the transmitter uses type $A_1$ in odd time slots, and type $A_2$ in even time slots. The use of two molecule types resembles MOSK but the difference is that the molecule type is not used for signalling. To convey information in each time slot, different diffusion rates are used (similar to CSK). Thus, to match each symbol to $b$ bits, $2^b$ different propagation rates are utilized in each time slot. As the molecule types are different in two subsequent time slots, (i) the previous symbol interference has almost disappeared (ii) the decision threshold of the current symbol is independent of the last decoded symbol.  Since the data is not encoded in the molecule types (but in the concentrations), the number of molecule types (and as a result the complexity), does not increase with $b$. We refer to this scheme as Molecular-Concentration Shift Keying (MOCSK).
%\vspace{-0.3cm}
\subsection{A Poisson model for molecular communication}

In this section we develop a model for a general molecular transmission. Later we use this model to compare probabilities of error of MOSK, CSK and MOCSK. It is assumed that the symbols are transmitted in equal time slots called symbol duration, $t_s$. Generally the signal is decoded based on the concentration of molecule types received during a time slot. We assume that in a time slot three sources contribute to the received molecules of a specific type: (i) molecules due to transmission in the current time slot (ii) the residue molecules due to transmission in the last time slot (the residue molecules from two or more earlier time slots are ignored) (iii) the molecules from other sources as the noise of environment. More precisely, each molecule either misses hitting the receiver, or hits the receiver in the current or next time slot. Let the hitting probabilities in the current and next time slot be $P_1$ and $P_2$, respectively ($P_1+P_2\leq 1$). We assume intra-molecule collisions have little effect on the molecule's movement.

For each molecule type, the transmitter has a storage. Every storage has an outlet whose size controls how many molecules can exit the storage to be diffused in the environment. The distribution of the number of molecules that exit from the storage in a time slot is Poisson. This follows from the fact that the number of molecules in the storage is large and probability of each passing through the outlet is small. We use $Poisson(X)$ to denote the number of molecules exiting the storage (and diffused in the environment). The parameter $X$ is called the diffusion rate and is determined by the size of the opening of the outlet. Parameter $X$ itself is a random variable that is decided by the transmitter based on the message it wishes to transmit.

Assume that the transmitter has chosen diffusion rates of $X_1$ and $X_2$ of a molecule type like $A$ in two consecutive time slots in a modulation scheme. Also, consider the environmet noise independent of the source signal as a Poisson noise with parameter $\lambda_0$. Considering hitting probabilities $P_1$ and $P_2$, and using the thinning property of Poisson and the fact that sum of independent Poisson rv's is itself a Poission, the number of type $A$ molecules received within the current time slot (denoted by $Y^{(A)}$) follows a Poisson distribution with parameter $\lambda$ %\begin{equation}
%Y^{(A)}\sim Poisson(\lambda).
%\end{equation}
where $\lambda=P_1 X_1+P_2 X_2+\lambda_0$. This Poisson model has advantages over the binomial model considered in \cite{K}: (i) The Poisson model is more realistic and practical in comparison with the binomial model. For instance, in binomial model the transmitter needs to count the number of molecules that it diffuses in comparison with the Poisson model in which the size of the storage's outlet is controlled (without counting the exact number of molecule passing through the outlet). (ii) The Poisson model is easier to work with analytically because of the nice properties of Poisson (e.g. thinning property, etc.).
%\vspace{-0.75cm}
\subsection{Decoding probabilities for CSK, MOSK and MOCSK}

In the CSK method, only one type of molecule for instance $A$ is used in all time slots. So, the decoding is made by comparing the number of received molecules, i.e. $Y^{(A)}$, with the thresholds determined based on the last decoded symbol.

In MOSK, as information is encoded in molecule type, if two subsequent symbols are the same, only one molecule type for instance $A_1$  is received, otherwise two molecule types for instance $A_1$ from the current symbol and $A_2$ type from the previous symbol are received. In the former case, the symbol is decoded correctly if the total sum of received molecules, i.e. $Y^{(A_1)}$, exceeds the threshold $\tau$. In the latter, decision is made based on $Y^{(A_1)}$, $Y^{(A_2)}$ and the threshold $\tau$.

In the proposed MOCSK scheme, as stated above, 2 different types of molecules in 2 subsequent time slots are used. As the receiver knows the type of the current symbol, it needs only to compare the number of the molecules of this type with the thresholds to decode the current symbol.
We now compute the average probability of error for CSK, MOSK and MOCSK.

As in \cite{K}, we only consider the class of simple symbol-by-symbol decoders that are practically appealing. Comparisons are made consistently for decoders in this class with the same transmission powers.

For binary CSK (BCSK) using molecules of type $A$, if the current input symbol is $s_c=0$ and the last transmitted and decoded symbols are $s_p$ and $\widehat{s}_p$ respectively, the probability of successful decoding is computed as\small
\small\begin{align*}
P_{\widehat{S}_c|S_c,S_p,\widehat{S}_p}(\widehat{s}_c=0|s_c=0,s_p,\widehat{s}_p )=P(Y^{(A)}<\tau_{\widehat{s}_p}),
\end{align*}\normalsize
where $\widehat{s}_c$ is the decoded current symbol and $\tau_{\widehat{s}_p}$ is the decision threshold corresponding to the value of $\widehat{s}_p$. Note that in this equation, $\widehat{s}_p$ is not necessarily equal to the previous transmitted symbol, i.e. $s_p$. In fact, the decision threshold for the current symbol is determined based on the previous decoded symbol regardless of being decoded correctly or not.

For binary MOSK (BMOSK), assume molecules type $A_1$ and $A_2$ are sent for 0 and 1, respectively. For the current transmitted symbol of $s_c=0$, probability of successful decoding for $s_p=1$ and $s_p=0$ are computed as follows:\\
\small
\begin{align*}
P_{\widehat{S}_c|S_c,S_p}(\widehat{s}_c=0|s_c=0,s_p=1)&=P(Y^{(A_1)}>\tau)P(Y^{(A_2)}\leq \tau),\\
P_{\widehat{S}_c|S_c,S_p}(\widehat{s}_c=0|s_c=0,s_p=0)&=P(Y^{(A_1)}>\tau).
\end{align*}
\normalsize
For the proposed binary MOCSK (BMOCSK), if the molecules of type $A$ is used in the current slot, the probability of successful decoding can be computed as follows:
\small\begin{align*}
P_{\widehat{S}_c|S_c}(\widehat{s}_c=0|s_c=0)=P(Y^{(A)}<\tau).
\end{align*}\normalsize
These conditional probabilities for quaternary and higher modulation level cases can be calculated similarly.
\section{Performance Analysis}
\subsection{Probability of error}
Let $E=\textbf{1}(\hat{S}_c\neq S_c)$ be the error event. For BCSK, the average probability of error is equal to $P_e=0.5P(E|S_c=0)+0.5P(E|S_c=1)$. As the decoding of the current symbol is dependent on both the last transmitted and decoded symbols in BCSK, we have:%The average probability of error is equal to $P_e=\sum_{s_c}{P(e,s_c)}$ where $P(e,s_c)$ is the probability that the transmitted symbol is $S_c=s_c$ and it is decoded erroneously at the receiver.
\small
\begin{align}
P(E|0) &=\sum_{s_p,\widehat{s}_p}{P_{S_p,\widehat{S}_p|S_c}(s_p,\widehat{s}_p|S_c=0)P(E|S_c=0,s_p,\widehat{s}_p)}\nonumber\\
&=\sum_{s_p,\widehat{s}_p}{P_{S_p}(s_p)P_{\widehat{S}_p|S_p}(\widehat{s}_p|s_p)P(E|S_c=0,s_p,\widehat{s}_p)},\nonumber\\
P(E|1)&=\sum_{s_p,\widehat{s}_p}{P_{S_p,\widehat{S}_p|S_c}(s_p,\widehat{s}_p|S_c=1)P(E|S_c=1,s_p,\widehat{s}_p)}\nonumber\\
&=\sum_{s_p,\widehat{s}_p}{P_{S_p}(s_p)P_{\widehat{S}_p|S_p}(\widehat{s}_p|s_p)P(E|S_c=1,s_p,\widehat{s}_p)}.\label{eq:a}
\end{align}\normalsize
%Then,\small
%\begin{align}
%P(e,s_c,s_p,\widehat{s}_p )=P(e|s_c,s_p,\widehat{s}_p)P(s_c,s_p,\widehat{s}_p )\label{eq:b}.
%\end{align}\normalsize
%As $s_c$ is independent of $s_p$ and $\widehat{s}_p$, we obtain:
%\small\begin{align}\label{eq:c}
%P(s_c,s_p,\widehat{s}_p)=P(s_c)P(s_p)P(\widehat{s}_p|s_p).
%\end{align}\normalsize
When $\widehat{s}_p \neq s_p$, $P_{\widehat{S}_p|S_p}(\widehat{s}_p|s_p)$ is the probability that the previous symbol $s_p$ is falsely decoded to $\widehat{s}_p$ and can be substituted by $P(E|s_p)$. So when $\widehat{s}_p=s_p$, $P_{\widehat{S}_p|S_p}(\widehat{s}_p|s_p)$ will be equal to $1-P(E|s_p)$. Therefore, $P(E|0)$ and $P(E|1)$ can be computed recursively from (\ref{eq:a}) which is a system of linear equations. For QCSK (and higher) levels, the probability of error can be computed similarly.% by solving a linear equation system with four (and more) unknown variales.

For the MOSK, the decoding of the current symbol is independent of the previous decoded symbol, i.e. $\widehat{s}_p$. Therefore, $P_e$ is obtained as
\small \begin{align*}
P_e=\sum_{s_c,s_p}{P(E|s_c,s_p)P_{S_c}(s_c )P_{S_p}(s_p)}.
\end{align*}\normalsize

For the proposed MOCSK, since the decoding of the current symbol is independent of both the transmitted and previous decoded symbols, $P_e$ is computed as follows
\small\begin{align*}
P_e=\sum_{s_c}{P_{S_c}(s_c)P(E|s_c)}.
\end{align*}\normalsize
\subsection{Lower bound on probability of error of molecular communication system}

In order to determine how well the modulation schemes considered above are, we examine a general molecular communication system with two molecule types $A_1$ and $A_2$ and a memory of only one symbol at the decoder (since we are considering the class of simple and practical symbol-by-symbol decoders only). Let $B_i, 1\leq i \leq k$ be the $i^{th}$ input symbol (assumed to be i.i.d. and uniform rv's over a set $\mathcal{B}$). The encoder uses $B_1,B_2,...,B_i$ to determine the diffusion rates $X_i^{(1)}$ and $X_i^{(2)}$ of molecules of type $A_1$ and $A_2$ respectively at the $i^{th}$ time slot. Thus the actual  number of molecules that are diffused would follow independent Poisson distributions with parameters $X_i^{(j)}$, $j=1,2$. Using the thinning property of Poisson, the distribution of the numbers of molecules of type $A_j$, $j=1,2,$ received at the receiver in the $i^{th}$ time slot is $Y_i^{(j)}\sim Poisson(P_1 X_i^{(j)}+P_2 X_{i-1}^{(j)}+\lambda_0)$, where $P_1$, $P_2$ and $\lambda_0$ are defined in section II-C. The current input symbol is decoded as $\widehat{B}_i$ based on the $Y_i^{(1)}, Y_i^{(2)}$ and $\widehat{B}_{i-1}$ (the previous decoded symbol). The average probability of error is:
\small\begin{align*}
P_e=\frac{1}{k}\sum_{i=1}^{k}{P(\widehat{B}_i\neq B_i)}.
\end{align*}\normalsize
Using the following lemma, a lower bound on $P_e$ is derived.\\
\textbf{Lemma 1.} $I(B_i,\widehat{B}_i )$ is bounded from above as follows:
\small\begin{align}\label{eq:lemma}
I(B_i;\widehat{B}_i)\leq I(Y^{(1)}_i;X^{(1)}_{i}|X^{(1)}_{i-1})+I(Y^{(2)}_i;X^{(2)}_{i}|X^{(2)}_{i-1}).
\end{align}\normalsize
\begin{proof}
The following chain of inequalities proves (\ref{eq:lemma}).\small
\small\begin{align*}
I(B_i;\widehat{B}_i)&\leq ^{a}I(B_i;B_{i-1},\widehat{B}_{i-1},Y^{(1)}_i,Y^{(2)}_i)\\
&\leq I(B_i;B_{i-1},\widehat{B}_{i-1},Y^{(1)}_i,Y^{(2)}_i,X^{(1)}_{i-1},X^{(2)}_{i-1})\\
&=^{b} I(B_i;B_{i-1},Y^{(1)}_i,Y^{(2)}_i,X^{(1)}_{i-1},X^{(2)}_{i-1})\\
%&=I(B_i;Y^{(1)}_i,Y^{(2)}_i|B_{i-1},X^{(1)}_{i-1},X^{(2)}_{i-1})\\
%&~~+I(B_i;B_{i-1},X^{(1)}_{i-1},X^{(2)}_{i-1})\\
&=^{c}I(B_i;Y^{(1)}_i,Y^{(2)}_i|B_{i-1},X^{(1)}_{i-1},X^{(2)}_{i-1})\\
&\leq I(X^{(1)}_{i},X^{(2)}_{i},B_i;Y^{(1)}_i,Y^{(2)}_i|B_{i-1},X^{(1)}_{i-1},X^{(2)}_{i-1})\\
&\leq I(X^{(1)}_{i},X^{(2)}_{i},B_i,B_{i-1};Y^{(1)}_i,Y^{(2)}_i|X^{(1)}_{i-1},X^{(2)}_{i-1})\\
&=I(X^{(1)}_{i},X^{(2)}_{i};Y^{(1)}_i,Y^{(2)}_i|X^{(1)}_{i-1},X^{(2)}_{i-1})\\
&\quad+I(B_i,B_{i-1};Y^{(1)}_i,Y^{(2)}_i|X^{(1)}_{i-1},X^{(2)}_{i-1},X^{(1)}_{i},X^{(2)}_{i}) \\
&=^{d}I(X^{(1)}_{i},X^{(2)}_{i};Y^{(1)}_i,Y^{(2)}_i|X^{(1)}_{i-1},X^{(2)}_{i-1})\\
&=H(Y^{(1)}_i,Y^{(2)}_i|X^{(1)}_{i-1},X^{(2)}_{i-1})\\
&\quad-H(Y^{(1)}_i,Y^{(2)}_i|X^{(1)}_{i-1},X^{(2)}_{i-1},X^{(1)}_{i},X^{(2)}_{i})\\
&\leq H(Y^{(1)}_i|X^{(1)}_{i-1},X^{(2)}_{i-1})+H(Y^{(2)}_i|X^{(1)}_{i-1},X^{(2)}_{i-1})\\
&\quad-H(Y^{(1)}_i|X^{(1)}_{i-1},X^{(1)}_{i})-H(Y^{(2)}_i|X^{(2)}_{i-1},X^{(2)}_{i})\\
&=I(Y^{(1)}_i;X^{(1)}_{i}|X^{(1)}_{i-1})+I(Y^{(2)}_i;X^{(2)}_{i}|X^{(2)}_{i-1}).
\end{align*}\normalsize

Inequality (a) holds because $\widehat{B}_i$ is function of $\widehat{B}_{i-1},Y_i^{(1)},Y_i^{(2)}$; (b) holds as $\widehat{B}_{i-1}$ does not provide further information about $B_i$ conditioned on $X_{i-1}^{(1)},X_{i-1}^{(2)}, B_{i-1}$. Equality (c) holds because $B_i$ is independent of previous message bits. (d) holds because of the Markov chain $(B_i,B_{i-1}) - X_i^{(1)} X_{i}^{(2)} X_{i-1}^{(1)} X_{i-1}^{(2)} - Y_i^{(1)} Y_i^{(2)}$.
% holds because the pair $X_i^{(1)}$ and $X_i^{(2)}$ are function of $B_i$ conditioned on $B_{i-1}$; (e) and (f)
\end{proof}
Using the well-known trick of taking a random index $Q$ uniformly distributed on $\{1,2,\cdots,k\}$ and independent of previously defined rv's, one can show $P_e=P(\hat{B}_Q\neq B_Q)$, and the statement of lemma 1 for random index $Q$ instead of fix index $i$. Considering constraints $0 \leq X_Q^{(j)} \leq r_{max}$ and $\mathbb{E}[X_Q^{(j)}]\leq r_{av}$ using the upper bounds on the discrete Poisson channels derived in \cite{LM} if $r_{av}/r_{max} \in[1/3,1]$ we have:
\small\begin{align}\label{eq:d}
I(X_{Q}^{(j)};Y_i^{(j)}|X_{Q-1}^{(j)})\leq\frac{1}{2}\log_{2}{P_1\gamma}+\frac{1}{2}\log_{2}{\frac{\pi e}{2}}+O_{P_1\gamma}(1),%\nonumber
\end{align}\normalsize
and if $r_{av}/r_{max}\in(0,1/3)$
\small\begin{align}\label{eq:e}
&I(X_{Q}^{(j)};Y_i^{(j)}|X_{Q-1}^{(j)})\\
&\leq 0.5\log_{2}{P_1 r_{max}}-(1-\frac{r_{av}}{r_{max}})\mu-0.5\log_{2}{2\pi e}\nonumber\\
&-\log_{2}{(0.5-\frac{r_{av} \mu}{ r_{max}})} +O_{P_1 r_{max}}(1),\nonumber
\end{align}\normalsize
where $O_{P_1 r_{max}}(1)$ tends to $0$ when $P_1 r_{max}$ tends to infinity and $\mu$ is the solution of $r_{av} /r_{max}=\frac{1}{2\mu}-\frac {e^{-\mu}}{{\sqrt{\mu \pi}erf(\sqrt{\mu})}}$.
Denoting the RHS of the upper bound inequality (\ref{eq:d}) or (\ref{eq:e}) by $G$, and using the upper bound in (\ref{eq:lemma}) we have,
\small\begin{align}
I(B_Q;\widehat{B}_Q)=H(B_Q)-H(B_Q|\widehat{B}_Q )\leq 2G.
\end{align}\normalsize
Using the Fano's inequality, $H(B_Q)=\log_{2}{|\mathcal{B}|}$ and the convexity of entropy function, we obtain:
\small\begin{align}\label{eq:LB}
H(P_e)+P_e \log_{2}{(|\mathcal{B}|-1)}\leq \log_{2}{|\mathcal{B}|}-2G,
\end{align}\normalsize
where $|\mathcal{B}|$ is the size of the alphabet of variable $B_i$. This inequality gives a lower bound on the average probability of error. Our analysis also extends to $M$ types of molecules in a straightforward way; it is omitted due to lack of space.
%\textbf{Remark 1.} If we extend the considered model to $M$ different types of molecules $A_1,A_2,\ldots,A_M$ , it is straight forward to see
%\small\begin{align}
%I(B_Q;\widehat{B}_Q)\leq \sum_{j=1}^{M} {I(X_Q^{(j)};Y_i^{(j)}|X_{Q-1}^{(j)})},
%\end{align}\normalsize
%and therefore, we have:
%\small\begin{align}
%H(P_e)+P_e \log_2 {(|B|-1)}\leq \log_2 {|B|}-MG.
%\end{align}\normalsize
\vspace{-0.2cm}
\section{Numerical Results}
For numerical evaluations, we have assumed the distance between transmitter and receiver and time slot duration are $d=16\mu m$ and $t_s=5.9$ sec. respectively and hence the probabilities of hitting as $P_1=0.22$ and $P_2=0.04$ (unless explicitly stated otherwise as in the description of two of the curves in Fig 1) \cite{KY}. Also, we have considered $\lambda_0=20$.

Fig. 1(Top) shows the plots of the probability of error versus the average transmission power (diffusion rate) per bit for binary and quaternary CSK, MOSK and proposed MOCSK modulations. This figure shows that BMOCSK substantially outperforms the BMOSK, with two types of molecules being employed in the two schemes. For the quaternary level, as expected, MOSK with more complicated transmitter and receivers employing four molecule types, yields a better performance compared to the proposed quaternary MOCSK (QMOCSK) that uses only two molecules types. The same results hold for the radio communication as well, where M-ary orthogonal modulation substantially outperforms M-ary amplitude modulation for $M\geq4$, at the cost of higher complexity. Also, Fig. 1(Top) shows that MOCSK outperforms CSK scheme in binary and higher levels. Further this improvement becomes more significant as we increase the hitting probability $P_2$. To illustrate this, probability of error of CSK has been plotted for $P_2=0.1$ in binary and quaternary levels in the Fig. 1(Top). Increasing $P_2$ results in more interference from the previous symbol, thus degrading the CSK performance. Generally speaking, as the decoding of the current symbol in the MOCSK is independent of both the previous transmitted and decoded symbols, this scheme outperforms the CSK and MOSK schemes in a fair comparison.

Fig. 1(Bottom) compares the probability of error of the proposed BMOCSK and the lower bound derived in (\ref{eq:LB}) versus the maximum transmission power. In the BMOCSK, $r_{av}/r_{max}=0.5$ and 2 molecule types are used. For a fair comparison, the lower bound has been computed for $M=2$, $r_{av}/r_{max}=0.5$ and $|\mathcal{B}|=2$. It is observed that the gap between the lower bound and the proposed scheme increases rapidly. To see how the lower bound changes with increasing $|\mathcal{B}|$, the lower bound has been also plotted for $|\mathcal{B}|=4,8,16$ and $M=2$.
\vspace{-0.5cm}
\begin{figure}%[t!]
\centering
\includegraphics[width=0.54\textwidth]{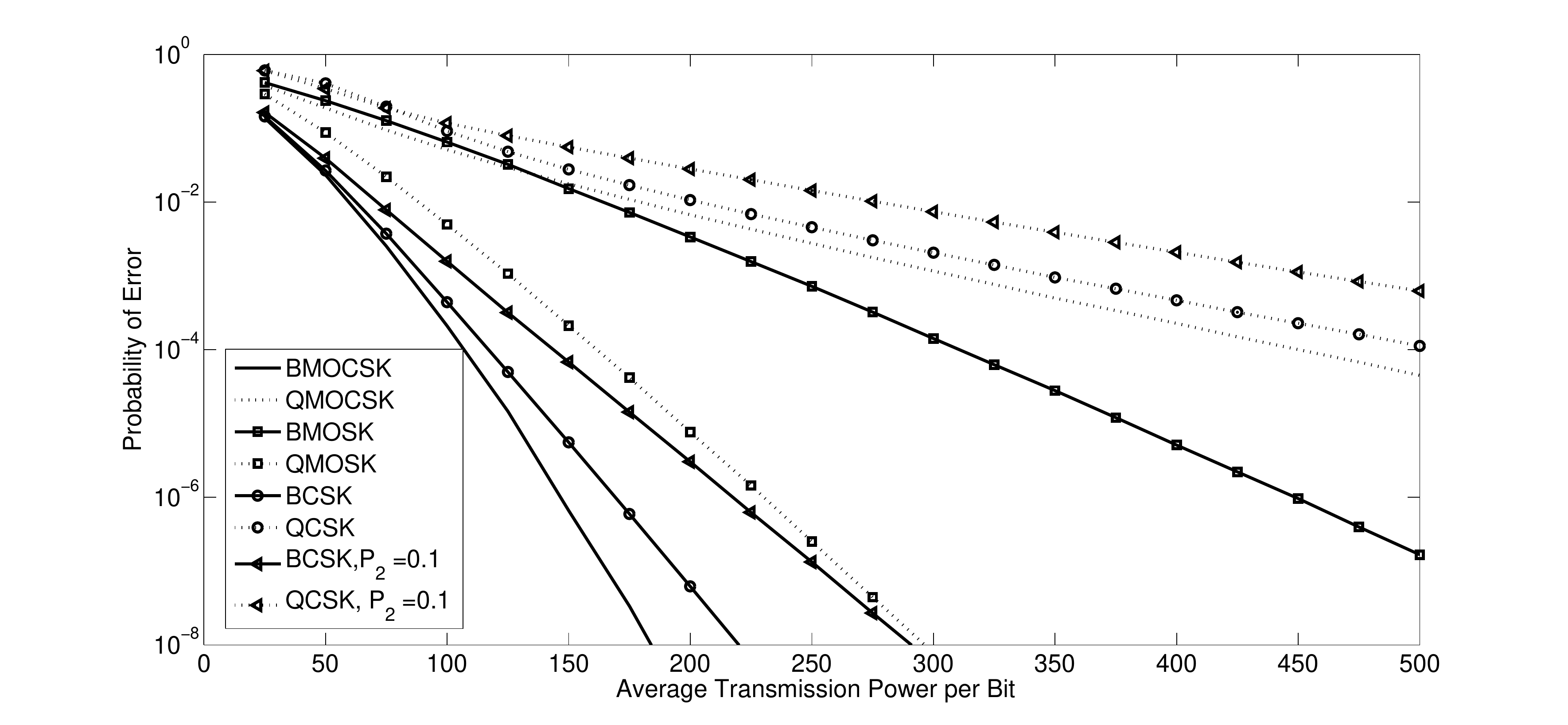}
\includegraphics[width=0.54\textwidth]{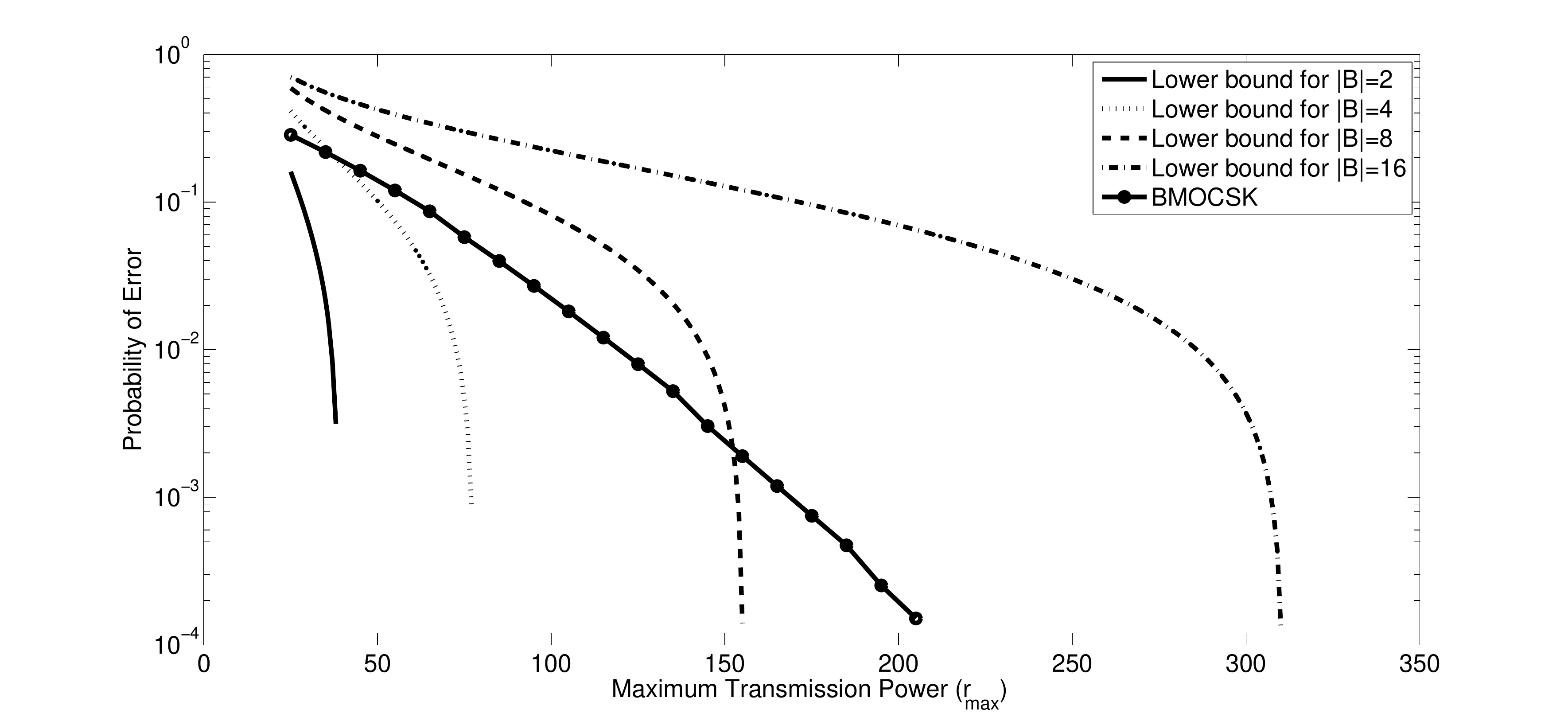}
\caption{Top: Average probability of error versus the average transmission power per bit. Bottom: Probability of error of the prposed QMOCSK and the lower bound versus the maximum transmission power.}
\vskip -0.7cm
\end{figure}
\ifCLASSOPTIONcaptionsoff
  \newpage
\fi

\end{document}